\documentclass[preprint,12pt]{elsarticle}

\usepackage{caption}
\usepackage{algorithm}
\usepackage{algorithmicx}
\usepackage{algpseudocode}

\usepackage{booktabs}
\usepackage{multirow}

\usepackage{color}

\usepackage{amsmath}

\newlength{\verticalcompensationlength}
\newcounter{verticalcompensationrows}
\newcommand{\verticalcompensation}[1]{%
  \setcounter{verticalcompensationrows}{#1}%
  \addtocounter{verticalcompensationrows}{-1}%
  \vspace*{-\value{verticalcompensationrows}\verticalcompensationlength}%
}

\newcommand{\multirowbt}[3]{%
  \multirow{#1}{#2}{\verticalcompensation{#1}#3}%
}

\usepackage{graphicx}
\usepackage{caption}
\usepackage{subcaption}

\usepackage{amssymb}

\journal{XXXXX}

\begin{document}

\begin{frontmatter}



\title{Multi-Linear Interactive Matrix Factorization}


\author[inst1]{Lu Yu}

\author[inst1]{Chuang Liu}

\author[inst1]{Zi-Ke Zhang\corref{cor1}}
\ead{zhangzike@gmail.com}
 \cortext[cor1]{Corresponding author.} 

\address[inst1]{Alibaba Research Center for Complexity Sciences, Hangzhou Normal University, Hangzhou 311121, PR China}

\begin{abstract}

Recommender systems, which can significantly help users find their interested items from the information era, has attracted an increasing attention from both the scientific and application society. One of the widest applied recommendation methods is the Matrix Factorization (MF). However, most of MF based approaches focus on the user-item rating matrix, but ignoring the ingredients which may have significant influence on users' preferences on items. In this paper, we propose a multi-linear interactive MF algorithm (MLIMF) to model the interactions between the users and each event associated with their final decisions. Our model considers not only the user-item rating information but also the pairwise interactions based on some empirically supported factors. In addition, we compared the proposed model with three typical other methods: user-based collaborative filtering (UCF), item-based collaborative filtering (ICF) and regularized MF (RMF). Experimental results on two real-world datasets, \emph{MovieLens} 1M and \emph{MovieLens} 100k, show that our method performs much better than other three methods in the accuracy of recommendation. This work may shed some light on the in-depth understanding of modeling user online behaviors and the consequent decisions.

\end{abstract}


\end{frontmatter}


\section{Introduction}
In recent years, the unprecedented proliferation of information has extremely changed our lifestyles. People all around the world are connected closely because of the daily basis millions of micro-blog posts, tweets and status updates of the social network. The popular online consumption is becoming an essential part of people's daily life, with the result that millions of e-commercial orders are generated per day. However, people are suffering from a serious and widely known problem: how to acquire quality recommendations from the numerous web service providers? Since the early work \cite{resnick1994grouplens} was published in 1990s, personalized $recommender\ systems$ (RS) \cite{lu2012recommender,bobadilla2013recommender} has been a thriving subfield of data mining to tackle this concern. 

In general, RS, serving as a special category of knowledge-based systems, attempts to automatically measure the relevance of user-user or item-item pairs, then delivers items to fit user's tastes via two basic strategies: $Content\ Based$ (CB) \cite{balabanovic1997fab} and $Collaborative\ Filtering$ (CF) \cite{herlocker1999algorithmic}. CB profiles items and users by extracting characteristic units from their content (e.g. demographic data, product information/description), and then identifies the matching-degree by comparing the corresponding profiles. However, due to the high cost to collect the necessary information about items and the lack of motivated users to share their personal data, CB fails to be the most popular recommendation approach. In contrast with CB, CF generates recommendations according to the structure of $virtual\ community$ \cite{hill1995recommending}. The virtual community is based on the underlying assumption that a group of people sharing similar characteristics in the past would also agree on their tastes in future. In addition, CF requires no domain knowledge and offers an alternative approach to reveal the latent patterns that are difficult to be captured by CB methods.


According to pioneering research, CF mainly contains two families: the $Neighborhood$ $Based$ $Models$ (NBMs) \cite{sarwar2001item,linden2003amazon} and the $Latent$ $Factor$ $Models$ (LFMs) \cite{hofmann2004latent, koren2009matrix}. NBMs namely outline the act of working together with neighbors. Here the term ``neighbor" does not only point to users, but also items, who share many characteristics in essence. Noteworthiness, $user$- and $item$- $based$ CF \cite{sarwar2001item, linden2003amazon} are two typical strategies to implement NBMs by measuring the likelihood of neighborhood between users or items with pre-defined similarity function. NBMs make predictions based on the known ratings involved with the active users'/items' neighbors. Comparatively, LFMs identify a couple of entities with the same dimensional feature vector inferred from the existing ratings, and straightly express the preference power with the dot-product of the corresponding feature vector pairs. On the basis of previous works, LFMs offer another idea to express various aspects or patterns of data, usually along with high accuracy and scalability.

As the most representative technique of LFM, Matrix Factorization (MF) results in numerous variants validated against the real data sets because of its high accuracy, scalability and expressive ability to capture various context factors (e.g. emotion, location, time). The earliest work of employing MF to implement CF was proposed by Sarwar \emph{et al.}, who conducted a  case study on the application of dimension reduction in CF with Singular Value Decomposition (SVD) method \cite{sarwar2000application}. Recently, Hofmann \cite{hofmann2004latent} reported on applying Latent Semantic Model to implementing LFM. At the beginning of Netflix Prize Competition \cite{bennett2007netflix} in 2006, Brandyn Webb detailed how the Regularized Matrix Factorization (RMF) \cite{webb2006rmf} helped his team rank in the third place under the pseudonym Simon Funk. Subsequently, several works \cite{koren2009matrix,koren2008factorization,koren2010collaborative,takacs2008investigation,takacs2008matrix} showed that RMF has played a significant role in the solution that won the Netflix Prize (NP). The attractive characteristics (e.g. methodological simplicity, easy incorporation of additional information, high accuracy) of RMF inspire many researchers to mine its potential from different aspects, such as \cite{ koren2009matrix, koren2008factorization,koren2010collaborative,paterek2007improving,takacs2009scalable,luo2012incremental, luo2013applying,zhang2014information} and so on.

As the aforementioned principles of LFMs, the standard RMF can be easily used to discover the latent relationship hidden in the interactions between two entities. In real life, people could take a number of factors into account before making a decision. However, it is difficult for RMF to integrate the interactions between users and the factors beyond items themselves. Though this challenge can be addressed by the Tensor Factorization (TF) \cite{karatzoglou2010multiverse}, the model complexity will grow exponentially with the number of contextual factors. Recently, Koren \cite{koren2008factorization} claimed a methodology to incorporate the RMF with neighborhood information. In addition, Koren \cite{koren2010collaborative} proposed a novel work on addressing temporal changes in user behaviors with matrix factorization models. Baltrunas \emph{et al.} \cite{baltrunas2011matrix} presented the context-aware matrix factorization, which models the interaction of the contextual factors with items. Ma \emph{et al.} \cite{ma2011recommender} extended the RMF by integrating the social regularization terms under the assumption that two users tend to have similar feature vectors if they are closely connected in social networks.

In this paper, we present a novel approach, namely the Multi-Linear Interactive Matrix Factorization (MLIMF), to model the interactions between users and the factors (e.g. emotions, locations, the time when the rating is given, movie genres, movie directors), which may have significant influence on the user's decision process. Generally, web systems could log multiple information correlated with customer's rating over a specific item. In our model, besides the interaction between the user-item pair, we represent the relationship between a specific user-factor pair in a same latent space. Then, through extending the standard RMF, we linearly integrate the total pairwise interactions together as the components of the customer's final rating decision to construct MLIMF. To clear the principles and application scenarios of MLIMF, we conduct two examples in two real datasets of Movielens with different size. Experimental results prove that, comparing with the standard RMF and other baseline algorithms, MLIMF could obtain better accuracy with linear complexity. The main contributions of this work include:


\begin{itemize}
   \item[-] In addition to the rating matrix, online users' rating decision could be probably influenced by some other factors. We propose that user could have a special interaction with each factor, and such pairwise relationship could be represented in a same latent space.
   \item[-] MLIMF, maintaining the principles and expressive scalability of MF, presents an alternative approach to take into account extra information based on the RMF. In fact, the key idea of MLIMF can be incorporated into other invariants of RMF.
   \item[-] Two application scenarios of MLIMF are given. First, we show that the extracted different feasible features from the training sample serving as the accessorial information which could have significant influence on the user's rating action. Then, we describe how to model the user's temporal dynamic preferences by integrating the time factor into MLIMF.
\end{itemize}

The remainder of this paper is organized as follows. Section 2 describes the preliminaries. In Section 3 we detail our proposed recommendation model. Section 4 gives two application scenarios. Experimental results are given in Section 5. Finally, Section 6 summarises this work and outlooks future work.

\section{Preliminaries}

The CF problem can be simply defined as generating personalized recommendations for a given user by seeking for a group of people or items with similar features from a finite data sample. In the area of CF, the user preferences over involved items are quantized into the user-item rating matrix $R_{|U|\times |I|}$, where $|U|$ and $|I|$ respectively denote the size of the given user set $U$ and item set $I$. Each entry $r$ at position ($u,\ i$) of $R\in \mathbb{R}^{|U|\times |I|}$, denoted by $r_{ui}$, presents the user $u$'s preference on item $i$, usually with high value expressing the strong relationship between the user-item pair. Typically, in terms of system's received specific feedback, $r_{ui}$ can be binary ($r_{ui}\in \{0, 1\}$), integers from a given range (e.g. $r_{ui}\in \{1, 2, 3, 4, 5\}$), or a continuous numerical interval (e.g. $r_{ui}\in [-5, 5]$). In practice, matrix R is usually very sparse and we can only observe a limited set, $X = \{(u_1, i_1, r_1), (u_2, i_2, r_2),..., (u_t, i_t, r_t)\}$, normally $|X|$ $\ll$ ${|U|\times |I|}$. Thereby, CF based recommendation tasks can be regarded as missing data estimation through the known user-item rating pairs.


%


\subsection{Regularized Matrix Factorization}
Among the huge amount of solutions to CF problem, RMF has been demonstrated to be superior to classic NBMs on the grand NP competition. Furthermore, numerous RMF variants are proposed to discuss the probable applications of MF and show their high efficiency and accuracy on several real rating datasets as well. Different from traditional NBMs, the goal of RMF is to approximate R by constructing two low-rank matrices. The basic principle of RMF is to map a pair of entities into the same low-dimension feature space. Thus each entity could be represented as a low-dimension feature vector. Taking the rating prediction problem as an example, let $f$ denote the dimension of the feature space. $P\in \mathbb{R}^{|U|\times f}$ denotes the user feature matrix where each row $p_u$ corresponds to a particular user $u$ and $Q\in \mathbb{R}^{|I|\times f}$ represents the item feature matrix where each row $q_i$ corresponds to a particular item $i$ (usually $f \ll$ min$(|U|, |I|)$). Then the rating approximation of user $u$ on item $i$ could be transformed as calculation of the dot-product of corresponding user-item feature vector pair,

\begin{equation}
\label{Eq.1}
\widehat{r}_{ui} = p_uq_i^T,
\end{equation}
where $\widehat{r}_{ui}$ is the estimate of $r_{ui}$. Usually, the values of parameters in $P$ and $Q$ can be learned from the training samples by applying the stochastic gradient decent (SGD) to optimize the objective function $J(P,\ Q)$:
\begin{equation}
\label{Eq.2}
\min _{U,I}\ J(P,\ Q) = \frac {1} {2}\sum _{u\in U}\sum _{i\in I}1(u,i)\left( r_{ui} - p_{u}q_{i}^T\right) ^{2} + \frac {\lambda} {2} (\left\| p_{u}\right\| _{F}^{2} + \left\| q_i\right\| _{F}^{2}),
\end{equation}
where $\left| \left| \cdot \right| \right| _{F}$ represents the  Frobenius norm. $1(u,i)$ is an indicator function and $1(u,i)=1$ if user $u$ rates item $i$, otherwise $1(u,i)=0$. The second term of Eq. (2) serves as the regularizing bulk for avoiding overfitting, meaning that the trained model has bad generalization for the new coming case. According to \cite{takacs2009scalable},  $\lambda$ is the weight parameter for the regularized term. As Eq. (\ref{Eq.2}) shows,  $J$ is a quadratic function with local minimum. Under the principles of SGD solver, the involved parameters of feature matrices, $P$ and $Q$, can be updated by moving in the opposite direction of the gradient for each training example. The optimized result could always be found after looping through all training samples for limited times, each of which is called a training epoch. In \cite{takacs2009scalable}, initializing each entry in $P$ and $Q$ with random values chosen from a pre-defined scale could speed up the convergence rate. For each training case, the algorithm generates estimation of $r_{ui}$ and computes the associated prediction error
\begin{equation}
  \label{Eq.3}
    e_{ui} = r_{ui} - p_uq_i^T.
\end{equation}
Then the corresponding feature vectors can be updated by the following rules:
\begin{equation}
\label{Eq.4}
\begin{aligned}
     \frac {\partial J}{\partial p_u} &= - e_{ui} \cdot q_i + \lambda p_u\\
     \frac {\partial J}{\partial q_i} &= - e_{ui} \cdot p_u + \lambda q_i\\
     p_u &\leftarrow p_u - \gamma\frac {\partial J}{\partial p_u}\\
     q_i &\leftarrow q_i - \gamma\frac {\partial J}{\partial q_i}
\end{aligned}
\end{equation}
where $\gamma$ denotes the learning rate. As the updating range of feature vectors goes up in proportion to learning rate, $\gamma$ is the key ingredient to influence not only the procedures of seeking for optimized parameters, but also the convergence rate for $J$. However, it's a tough job to set a suitable value to $\gamma$ in real application of SGD. Though Luo \emph{et al.} \cite{luo2013applying} recently tried to deal with the dilemma of learning rate tuning through learning rate adaptation, it's still lack of uniformed policy to set the value of $\gamma$. Like most of proposed MF-based works, we regard $\gamma$ as an empirical parameter, adapting to different data sets. Besides setting suitable value to $\gamma$ and $\lambda$, Tak{\'a}cs \emph{et al.} \cite{takacs2009scalable} suggested that an early stopping criterion is necessary for avoiding overfitting. Usually, we can stop training the model until the evaluation metric on testing set does not improve any more, or fluctuates to a converged value.

The basic RMF has been proved to be highly accurate and scalable. However, many users offer very few ratings, which makes it difficult to identify their tastes. Fortunately, the MF approach is flexible in dealing with this problem by incorporating additional sources of information beyond the user-item rating matrix. In real applications, besides the users' explicit ratings, RS could easily capture the implicit feedback (e.g. browsing or purchases history, time effects) and social relationships to deeply analyze user preferences. To utilize the implicit feedback, Koren \cite{koren2008factorization} presented an alternative approach, named SVD++, to incorporate implicit information and user attributes into MF model. Jamali \emph{et al.} \cite{jamali2010matrix} reported the effect of trust propagation for recommendation in social networks. These works regard extra sources as significant elements that extensively influence the interactions between users and items. Alternatively, in this paper, we suppose that people tend to weight each extra factor into the final rating decision. This weighting-process just seems that in the sports competition, judgers could firstly measure athlete's performance in various aspects, then give the final score by synthetically taking into account the weights of all involved elements. Thereby, we model user's process of weighting each factor as an unique interactive result based on the RMF model. The final estimation of user's preferences on item is made by a simple linear combination of the interactive weight involved with each factor.

\subsection{Temporal Dynamic Matrix Factorization}

The aforementioned applications of MF models can not adapt to dynamic customer preferences. Usually concepts (e.g. customer preferences, item popularity, social structure) involved with data are changing over time, and models should distinguish short term effects from the longer term trends that reflect the intrinsic patterns of the data. Nonetheless, temporal changes in data bring unique challenges. With the detailed analysis, the possibility of modeling time effects on the performance of CF has been demonstrated by the recent works. Lathia \emph{et al.} \cite{lathia2008knn} analysed the evolution of retrieved characteristics over time and gave insightful explanations why certain CF similarity measures outperform others. In Ref. \cite{koren2010collaborative}, Koren suggested that temporal modeling should be a predominant factor in building RS, and proposed timeSVD++ to model the temporal drifting concepts. Therefore, according to previous researches \cite{koren2010collaborative, xiang2009time}, incorporating time effects into MF models has become a comparative mature topic.

Note that, those models include day-specific parameters for each user, which limits the feasibility for predicting their future ratings. In this paper, due to the pioneering discussions of the possible types of time effects, we attempt to model time effect as a decision factor for users to express the idea of our proposed MLIMF in the following section.


\section{Recommendation With MLIMF }
In this section, we will describe the definition of our focused problem, extending RMF to model the interactions between users and decision factors. In order to offer a better understanding, we conduct two applications of the proposed model.

\subsection{Problem Definition}
\begin{figure*}[htb]
\centering
  \includegraphics[scale = 0.16]{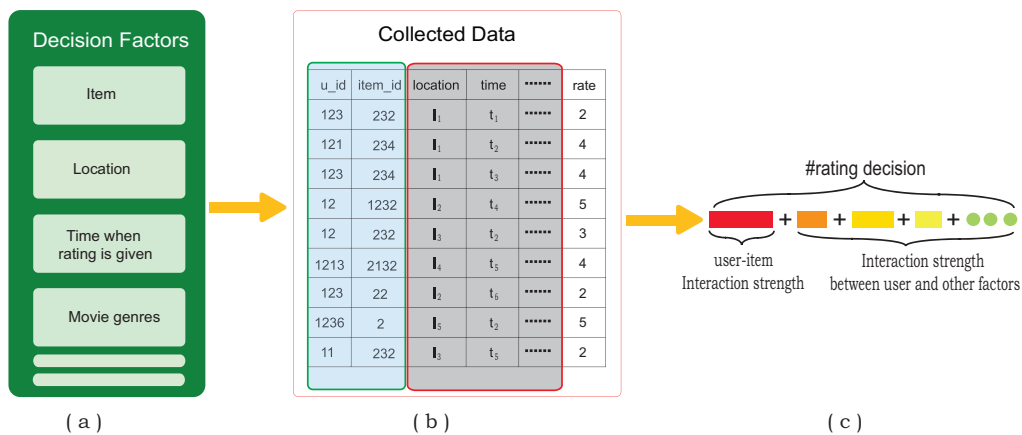}
  \caption{Illustration of each event along with the path to the last rating decision. (a) describes the possible factors affecting users' preferences over items. (b) presents the collected data after normalizing decision factors. (c) offers an intuitive understanding on how users weight each factor into the last rating decision.}
  \label{Fig:fig1}
\end{figure*}
With the purpose of both improving user experience and enhancing competitive power, electronic retailers and content providers would offer adequate information for a vast selection of products, which increases opportunities to meet customers' various personalized needs and tastes. Certainly, customers profit from the abundant data, which provide enough evidences to demonstrate the quality of involved products. As the figure \ref{Fig:fig1} shows, the final decision for purchasing a product may be influenced by many factors, such as emotion, seasonal discount, comments on the product and so on, and the impact of each on users is unbalanced. For example, as a big fan of $Star\ Wars$, user $u$ prefers to pay for another exemplary analogical movie (e.g. THX 1138), directed by $George\ Lucas$ (an American director famous for the series of $Star\ Wars$). However, such delicate information is not always available. Alternatively, extra sources associated with an active customer's ratings can always be captured by RS. In this paper, we model such effect under the assumption that user $u$'s preference to item $i$ can be parted into limited weighted components, each of which denotes the significance of involved factor in the final decision of $u$.

\subsection{Principles of MLIMF}
Based on the framework of RMF, the interaction between user $u$ and specific factor $j$ can be represented as the dot-product of corresponding low-rank feature vectors. Thereby, $\widehat{r}_{ui}$ can be modified as the following:
\begin{equation}
\widehat{r}_{ui} = p_uq_i^T + \sum _{j\in D}\sum _{d_j\in D_j}1(u,i,d_{j})p_{ud_{j}}q_{d_{j}}^T,
\label{Eq.5}
\end{equation}
where $D$ denotes the decision factor set. In Eq. (\ref{Eq.5}), the first term presents user $u$'s preference on item $i$, and the bulk behind $\sum$ notion denotes the interactions between user $u$ and possible decision factor $j$, which is always a categorical attribute with a set, denoted as $D_j$, of limited amount of values. It's noted that the indicator function $1(u,i,d_{j})$ is set as 1 if user $u$ focuses on the specific value of factor $j$, denoted as $d_j \in D_j$, when giving rating on item $i$, otherwise $1(u,i,d_{j})$ is set as 0. In other words, $d_j$ indicates the specific contextual information when user $u$ give his/her rate to item $i$. For example, a user who has ever rated ``5 stars" on a $Jackie\ Chan$'s movie $Rush\ Hour1$, might give higher weight to another movie played or directed by him. Here, in order to model the relationship between users and extra information, a new set of decision factor feature vectors are necessary, where $d_j$ is associated with feature vector $q_{d_{j}} \in \mathbb{R}^{f_{D_j}}$. $f_{D_j}$ denotes feature dimension parameter for decision factor $D_j$. Correspondingly, we define a new set of user feature vectors, where user $u$ involved with factor $j$  is associated with $p_{ud_{j}}\in \mathbb{R}^{f_{D_j}}$. Then the objective function $J$ is modified as follow:

\begin{equation}
\label{NCost}
\begin{aligned}
     \displaystyle\min _{U,I,D}\ J=&\ \displaystyle\frac {1} {2}\sum _{u\in U}\sum _{i\in I}1(u,i)( r_{ui} - p_{u}q_{i}^T - \sum _{j\in D}\sum _{d_j\in D_j}1(u,i,d_{j})p_{ud_{j}}q_{d_{j}}^T) ^{2}\\
     &{\displaystyle + \frac {\lambda} {2} (\sum _{u\in U}\sum _{j\in D}\sum _{d_{j}\in D_{j}}\left\| p_{ud_{j}}\right\| _{F}^{2} + \sum _{j\in D}\sum _{d_{j}\in D_{j}}\left\| q_{d_{j}}\right\| _{F}^{2})}\\
     &{\displaystyle + \frac {\lambda} {2} \left(\left\| p_{u}\right\| _{F}^{2} + \left\| q_i\right\| _{F}^{2} \right)}
\end{aligned}
,
\end{equation}
where the relationship between $1(u,i,d_j)$ and $1(u,i)$ has been illustrated in Figure 1b, where the blue rectangle represents the fact that user $u$ has selected item $i$, and the gray rectangle further points out the underlying decision factors $d_j$ when this rating record is generated. To make it easy to understand their relationship, we could intuitively review users` rating procedures. We firstly should make sure whether user u has shown his/her preference to item i, which results in adding $1(u, i)$ outside the bracket in Equation (6). Then interactions between users and underlying factors that users might take into consideration when giving rate to item i, are weighted into the final rating decision. It directly contributes to using $1(u, i, d_j)$ to depict such idea, where $d_j$ only represents the specific value of related decision factor to the contemporary rating record.

Eq. (\ref{NCost}) is more complicated than Eq. (\ref{Eq.2}) after including the regularized terms for feature vectors of extra sources. However, under the framework of SGD solver, the training parameters can be learned in linear time. Analogous with Eq. (\ref{Eq.4}), we calculate the gradients of the involved parameters with the following rules:
\begin{equation}
\label{DGradient}
   \begin{aligned}
     \frac {\partial J}{\partial p_u} &= - e_{ui} \cdot q_i + \lambda p_u\\
     \frac {\partial J}{\partial q_i} &= - e_{ui} \cdot p_u + \lambda q_i\\
     \frac {\partial J}{\partial p_{ud_{j}}} &= - e_{ui} \cdot q_{d_{j}} + \lambda p_{ud_{j}}\\
     \frac {\partial J}{\partial q_{d_{j}}} &= - e_{ui} \cdot p_{ud_{j}} + \lambda q_{d_{j}}
   \end{aligned}
   .
\end{equation}
Then for each training example with format $[u$, $i$, $d_1$, $...$, $d_{|D|}]$, the updating rule for model parameters is formulated by:
\begin{equation}
\label{EUpdating}
   \begin{aligned}
     p_u &\leftarrow p_u + \gamma(e_{ui} \cdot q_i - \lambda p_u)\\
     q_i &\leftarrow q_i + \gamma(e_{ui} \cdot p_u - \lambda q_i)\\
     p_{ud_{j}} &\leftarrow p_{ud_{j}} + \eta (e_{ui} \cdot q_{d_{j}} - \lambda p_{ud_{j}})\\
     q_{d_{j}} &\leftarrow q_{d_{j}} + \eta (e_{ui} \cdot p_{ud_{j}} - \lambda q_{d_{j}})
   \end{aligned}
   .
\end{equation}

By combing Eq. (\ref{DGradient}) and Eq. (\ref{EUpdating}), the values of model parameters can be efficiently learned after several epoches. However, in real applications, it is not easy to incorporate additional sources into MLIMF, due to the lack of motivated users to share their personalized tastes on each event along with the path to the last rating decision.

Nevertheless, online service providers still carefully polish the design of the software systems to capture more details for better understanding user behaviors, which plays an essential role in offering personalized and novel services to users, as well as enhancing the company reputation and competitive strength. In fact, the logged data in the database offers a highly possible approach to model users' rating action. Thereby, the issue of modeling the users' procedure of weighting decision factors becomes a problem on how to weight the modified and extracted probable features that might influence the users' rating decision for a particular item. To deeply clarify the principles of MLIMF, we conduct two possible applications on two real data sets.

\begin{algorithm}
\caption{Training algorithm for MLIMF}
\begin{algorithmic}[1] 
    \Require $T$: Train Set, $V$: Validation
    \Ensure Learned feature parameters
    \State Randomly initialize feature parameters from distribution $\mathcal{N}(0, 0.02)$.
    \Repeat
         \For{each observed sample [u, i, $d_1$, $d_j$ \ldots $d_{|D|}$, $r_{ui}$] in $T$}
            \State compute $\hat{r}_{ui}$ with Eq. (\ref{Eq.5});
            \State $e_{ui}$ = $r_{ui}$ - $\hat{r}_{ui}$;
            \State $p_u \leftarrow p_u + \gamma(e_{ui} \cdot q_i - \lambda p_u)$
            \State $q_i \leftarrow q_i + \gamma(e_{ui} \cdot p_u - \lambda q_i)$
            \State $p_{ud_{j}} \leftarrow p_{ud_{j}} + \eta (e_{ui} \cdot q_{d_{j}} - \lambda p_{ud_{j}})$
            \State $q_{d_{j}} \leftarrow q_{d_{j}} + \eta (e_{ui} \cdot p_{ud_{j}} - \lambda q_{d_{j}})$
         \EndFor
         \State Calculate the RMSE on V;

    \Until RMSE on V does not improve.
\end{algorithmic}
\end{algorithm}

\subsubsection{Recommendation with Extracted Features}
Before carrying on the data mining methods, pre-processing raw data sources can give an insight into the hidden interesting patterns. This subsection highlights the first application of MLIMF on two $Movielens$\footnote{Movielens is an online website with ultimate goal to gather research data on personalized recommendations systems. http://movielens.umn.edu/} data sets\footnote{http://grouplens.org/datasets/movielens/}:

\begin{itemize}
   \item[-] MovieLens 100k (ML100k) is collected by the GroupLens Research Project at the University of Minnesota via the MovieLens web site. ML100k contains 100,000 ratings (1-5) from 943 users on 1682 movies during the seven-month period from September 19th, 1997 to April 22nd, 1998. Each user has rated at least 20 movies. The density of the rating matrix in ML100k is 6.30\%. In addition to movie ratings, ML100k also provides various information on individual films, such as a group of genres and release date, which are used to increase the film recommendation system's accuracy.
   \item[-] MovieLens 1M (ML1m)  is another collected data set on Movielens web site, which contains 1,000,209 anonymous ratings (1-5) of approximately 3,900 movies and 6,040 MovieLens users who joined MovieLens in 2000. The density of the rating matrix in ML1m is 4.25\%. Like the ML100k, ML1m provides the same information on individual films.
\end{itemize}

Obviously, the published MovieLens data sets only collect one type of explicit feedback (users' ratings on movies), which simplifies users' decision  procedure of giving ratings to movies. In fact, the rating on a specific movie reflects a user's personalized attitude to the corresponding information of films. Although users do not explicitly express their viewpoints on each piece of movie information, the accumulative rating behaviors may imply interesting patterns. The core idea of CF is to utilize the accumulative data to estimate user preference on items under the assumption that a group of close neighbors with similar tastes could help each other rate objects. Based on the available information in the data, we can distill several feasible ingredients, which might affect user's rating decision. Then we incorporate those ingredients into the proposed MLIMF to model users' rating behaviors. By combing the previous Eqs. (\ref{DGradient}-\ref{EUpdating}), we can automatically learn the strength of interactions between user and those ingredients from the given data.

In this example, it's noted that besides the rating on item $i\in I$ by user $u\in U$ at time $t\in \mathbb{R}$, ML100k and ML1m also offer information on each film, denoted as set $S$, where
\begin{equation*}
\begin{aligned}
    S = \{&(A, Titanic, 2010.01.02, 5), (A, Star\ Wars, 2010.04.01, 1),\\
    &(B, Star\ Wars, 2009.05.04, 4), (C, Time\ Tracers, 2010.04.01, 4)\}.
\end{aligned}
\end{equation*}
Both ML100k and ML1m contain 19 types of genres and release date for each individual film and every movie can have multiple genres, namely $genre\ group$, which can highly reflect the users' tastes. The size of genre group describes individual bias on multi-genre movies. Through the feasible transformation on the observed data, we extract three additional features, release date ($RD$), genre group ($GG$), the size of corresponding genre group ($GS$). Consequently, each piece of observed data can be denoted as $(u,i,RD, GG, GS, r_{ui})$. Let the modified observed data $\hat{S}$ be:
\begin{equation*}
\begin{aligned}
    \hat{S} = \{&(A, Titanic, 1997, G1, 2, 5), (A, Star\ Wars, 1999, G2, 4, 1),\\
    &(B, Star\ Wars, 1999, G2, 4, 4), (C, Time\ Tracers, 1995, G3,  4, 4)\}\\
    G1 = &\{Drama,\ Romance\}\\
    G2 = &\{Action,\ Adventure,\ Fantasy,\ Science\ Fiction\}\\
    G3 = &\{Action,\ Adventure,\ Science\ Fiction\}
\end{aligned}
\end{equation*}
Usually we denote the user-item pairwise relationship as the rating matrix $R$. Thus the interactions between user and a specific factor $j$ can be analogically denoted as user-factor matrix $R_j$, each entry of which is a binary indicator, which is set as 1 if user $u$ is associated with factor $j$, otherwise set as 0.

\begin{figure*}[htb]
\centering
\begin{tabular}{cc}
  \includegraphics[scale = 0.29]{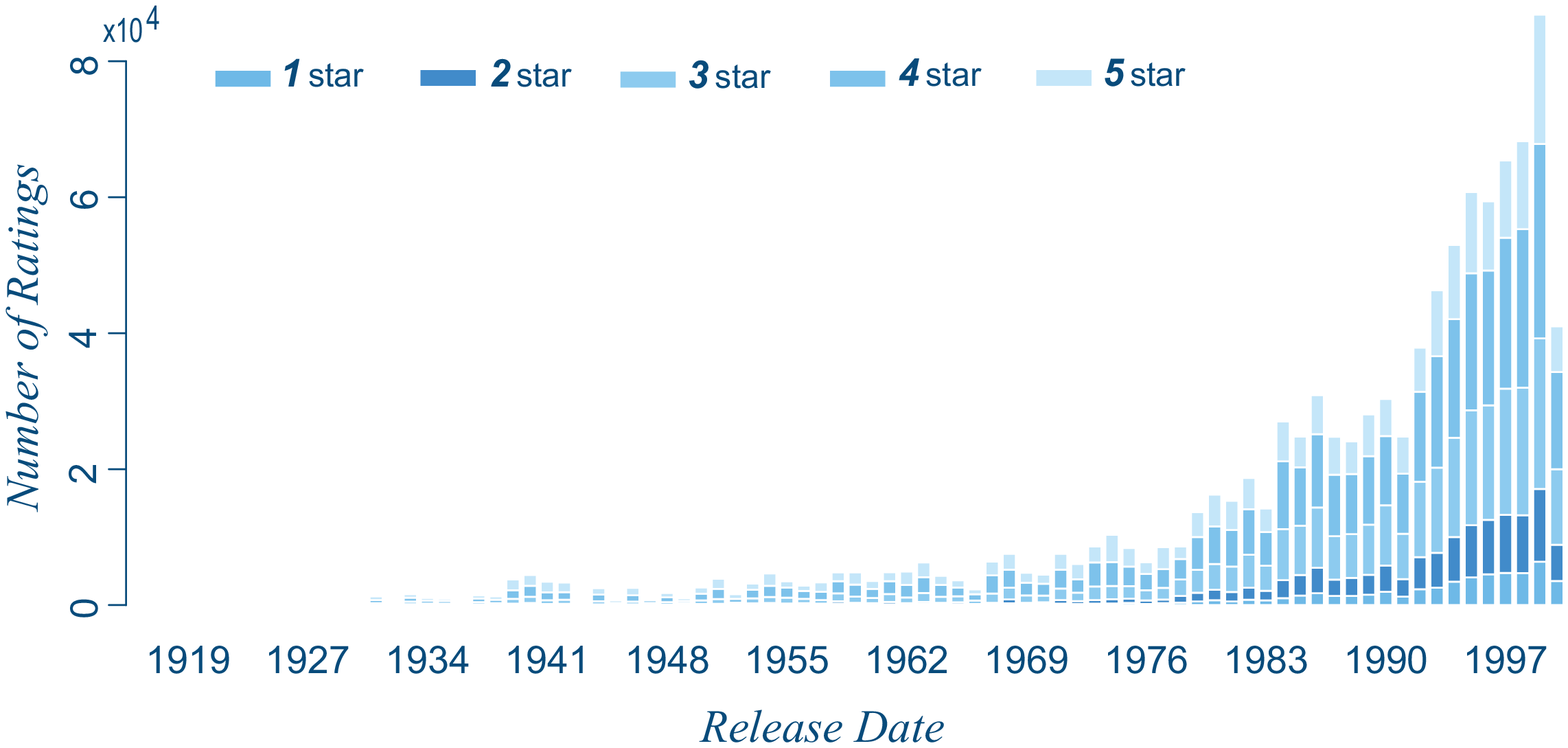}&
  \includegraphics[scale = 0.29]{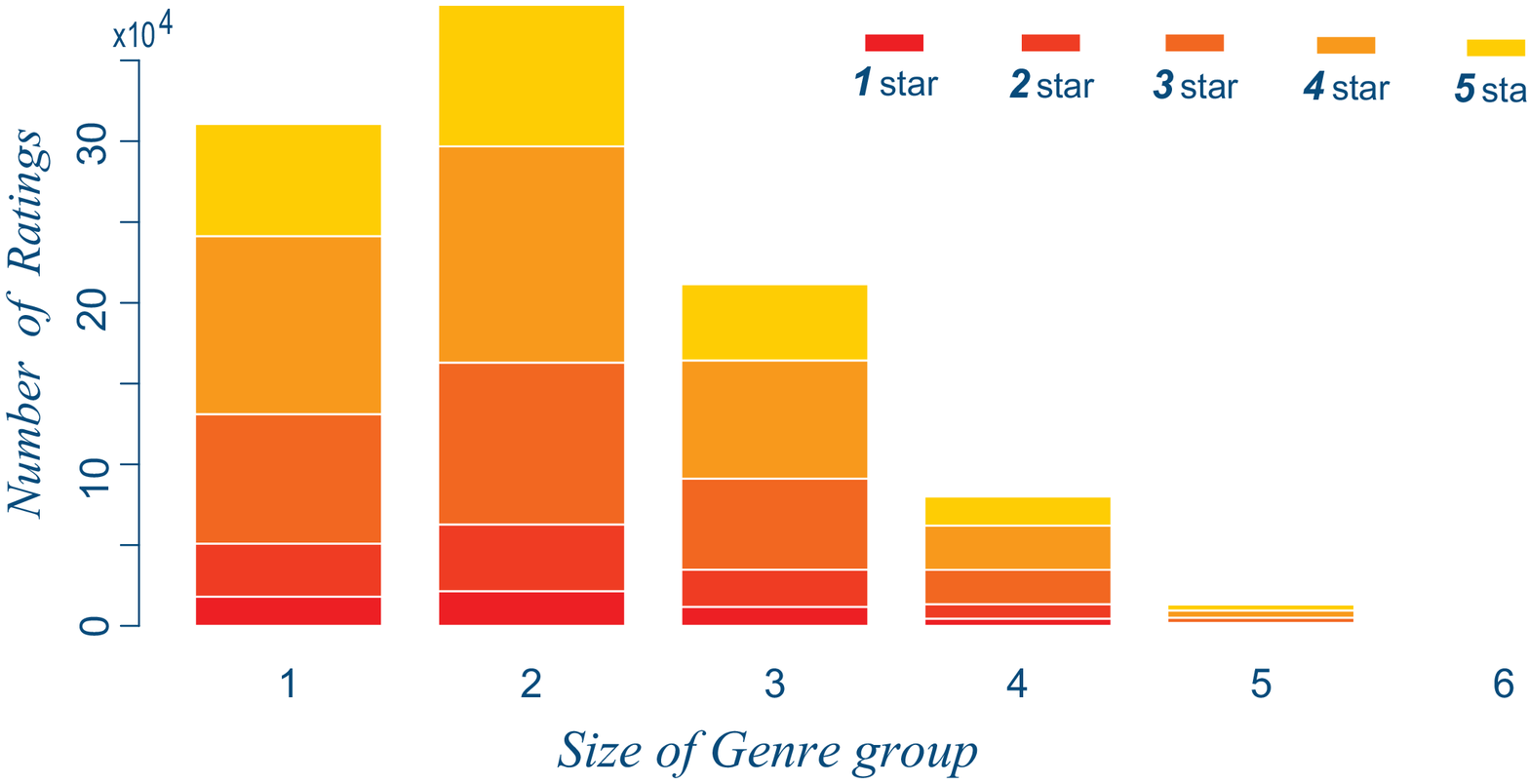}\\
  (a) & (b)
\end{tabular}
\caption{ Rating distribution for possible factors extracted from ML1m: (a) $Release\ Date$, (b) $Size\ of\ Genre\ group$. }
\label{Fig:Rates}
\end{figure*}

\begin{figure*}[htb]
\centering
\begin{tabular}{ccc}
  \includegraphics[scale = 0.25]{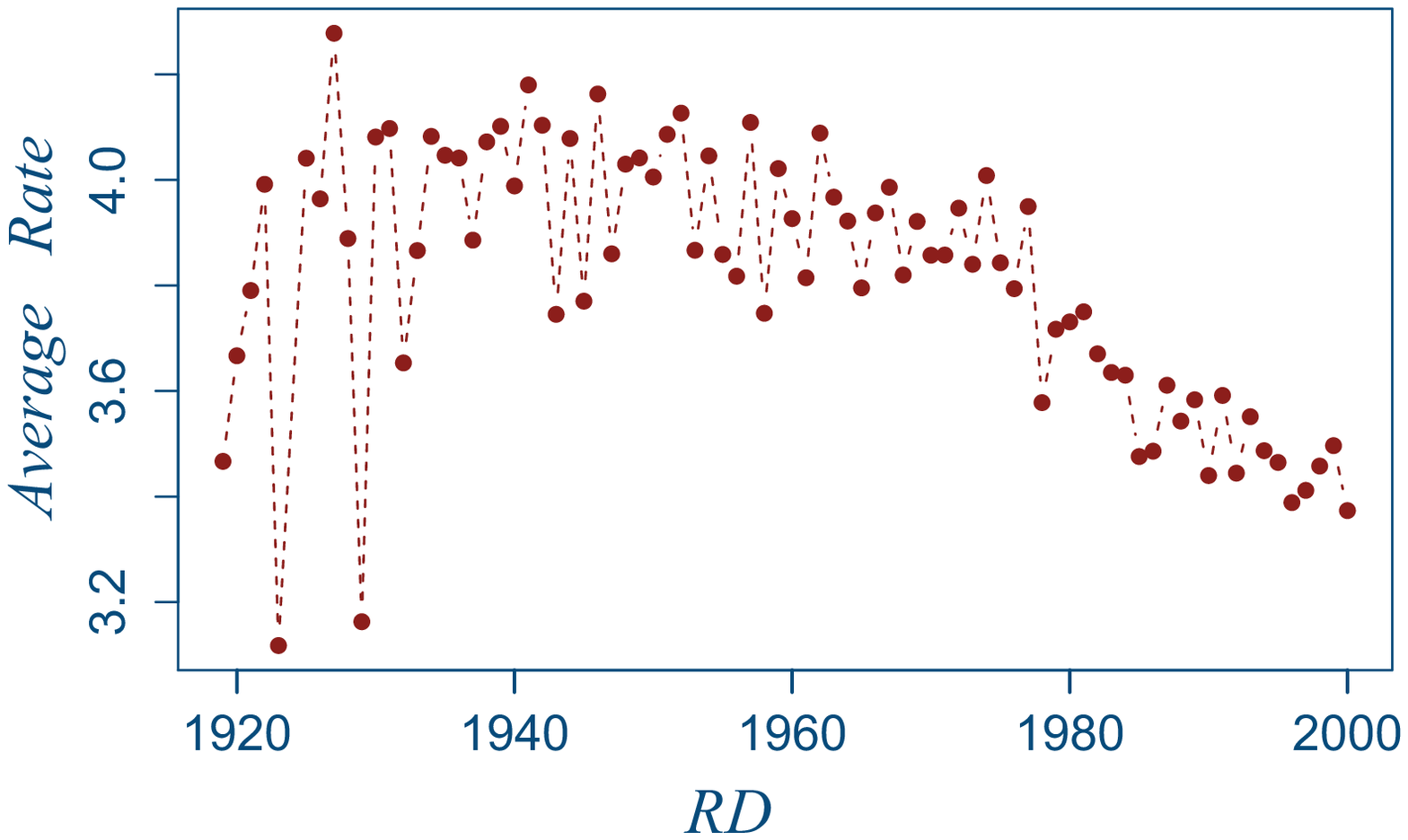}
  &\includegraphics[scale = 0.25]{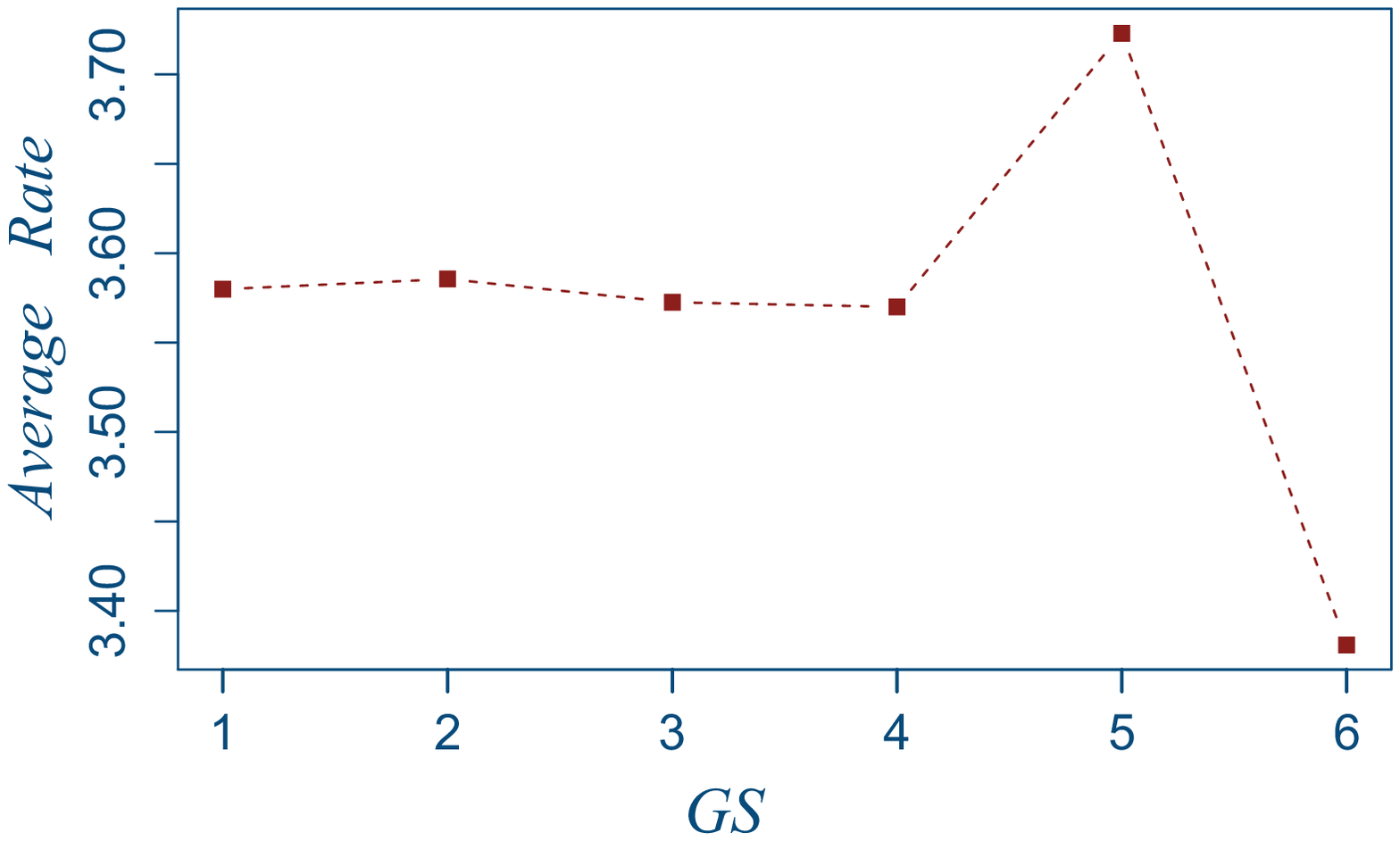}
  &\includegraphics[scale = 0.25]{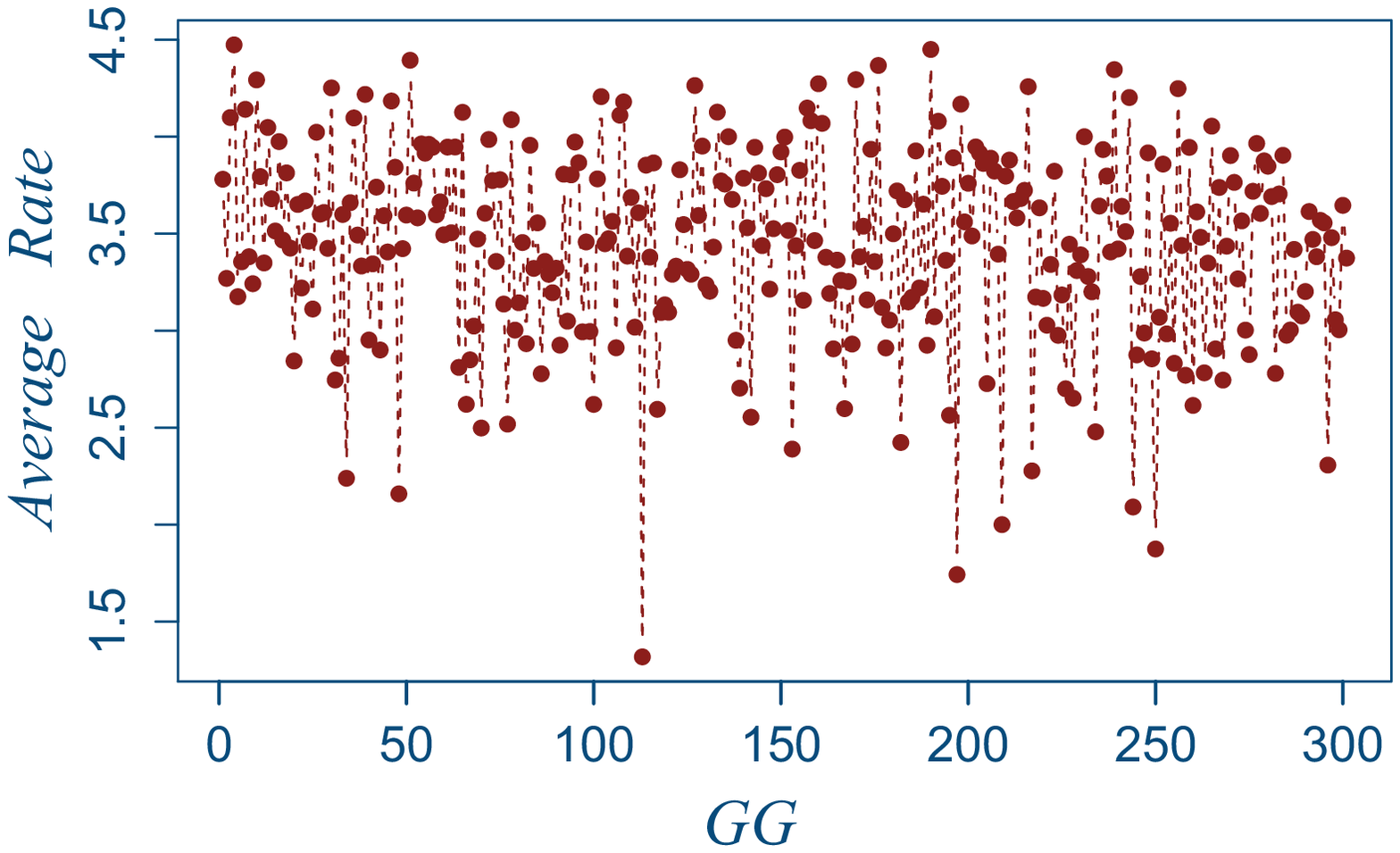}\\
  (a) & (b) & (c)
\end{tabular}
\caption{ Average rate distribution for possible factors extracted from ML1m: (a) $Release\ Date$, (b) $Size\ of\ Genre\ group$, (c) $Genre\ Group$.}
\label{Fig:Aver}
\end{figure*}

In real applications, users only directly give ratings to movies. However, facing different contextual environment, users might have a specific rating pattern for each factor. Figure \ref{Fig:Rates} and figure \ref{Fig:Aver} show the evolution of rating distribution for two factors extracted from the data. The distribution for factor $RD$, described in figure \ref{Fig:Rates}a, shows that people prefer to giving ratings to recent released movies. And figure \ref{Fig:Aver}a offers an evidence to demonstrate that people tend to give strict ratings as the release date grows. Interestingly, figure \ref{Fig:Aver}b shows that movies with 5 genres receive higher rating on average. Figure \ref{Fig:Aver}c depicts that the ratings given on movies could evolve with the movie genres.

For the prediction task of MLIM using the data like $\hat{S}$, the estimate value for an uncollected item $i$ of user $u$ can be formulated by:
\begin{equation}
\label{Eq.8}
\widehat{r}_{ui} = \underbrace{p_uq_i^T} _{user\ \#\ item} +\ \underbrace{p_{ud_1}q_{d_1}^T + p_{ud_2}q_{d_2}^T+ p_{ud_3}q_{d_3}^T} _{user\ \#\ decision\ factors},
\end{equation}

\subsubsection{Temporal Recommendation}
According to the pioneering research \cite{koren2010collaborative}, Koren suggests that modeling time effects is essential for building RS. Customer preferences for items are constantly changing over time. The product popularity also evolves over time when new selection emerges. Within the complex systems intersecting multiple customers and items, various characteristics are drifting over time, while many of them often are too delicate to be explored with a few data instances. In Refs. \cite{koren2010collaborative, xiang2009time}, they model time changes at the level of each individual, leading to modify day-specific variables. However, day-specific parameters are associated with certain past time points, which turns out to fail in predicting the changes in the future. In e-commerce systems, user feedbacks are constantly generated and vary at different time points. Analyzing such data brings unique challenges on finding the right balance between avoiding temporary effects that tinily affect the future behaviors, while capturing the long term trends that reflect users' regular patterns.

\begin{figure*}[htb]
\centering
\begin{tabular}{cc}
  \includegraphics[scale = 0.36]{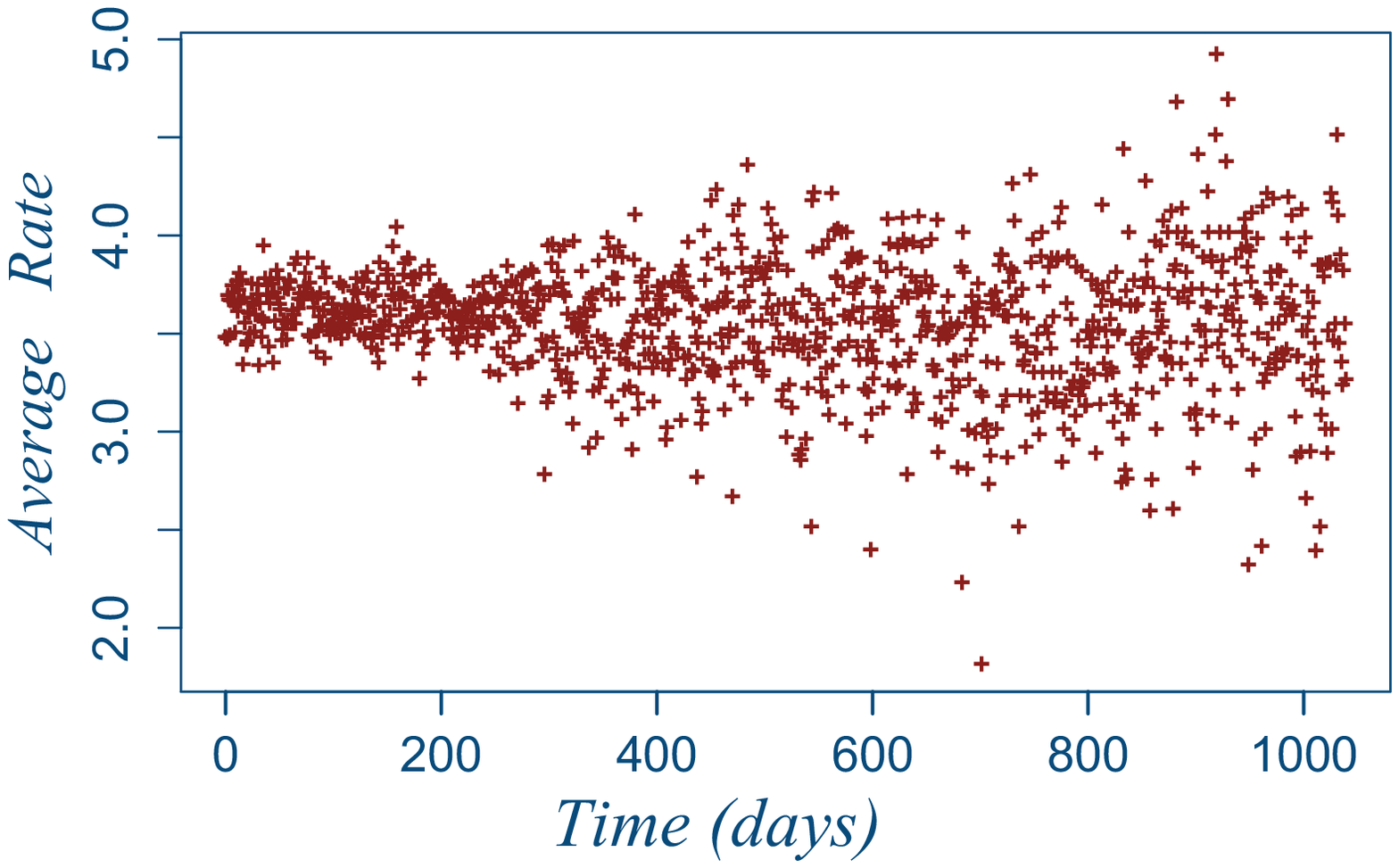}
  &\includegraphics[scale = 0.36]{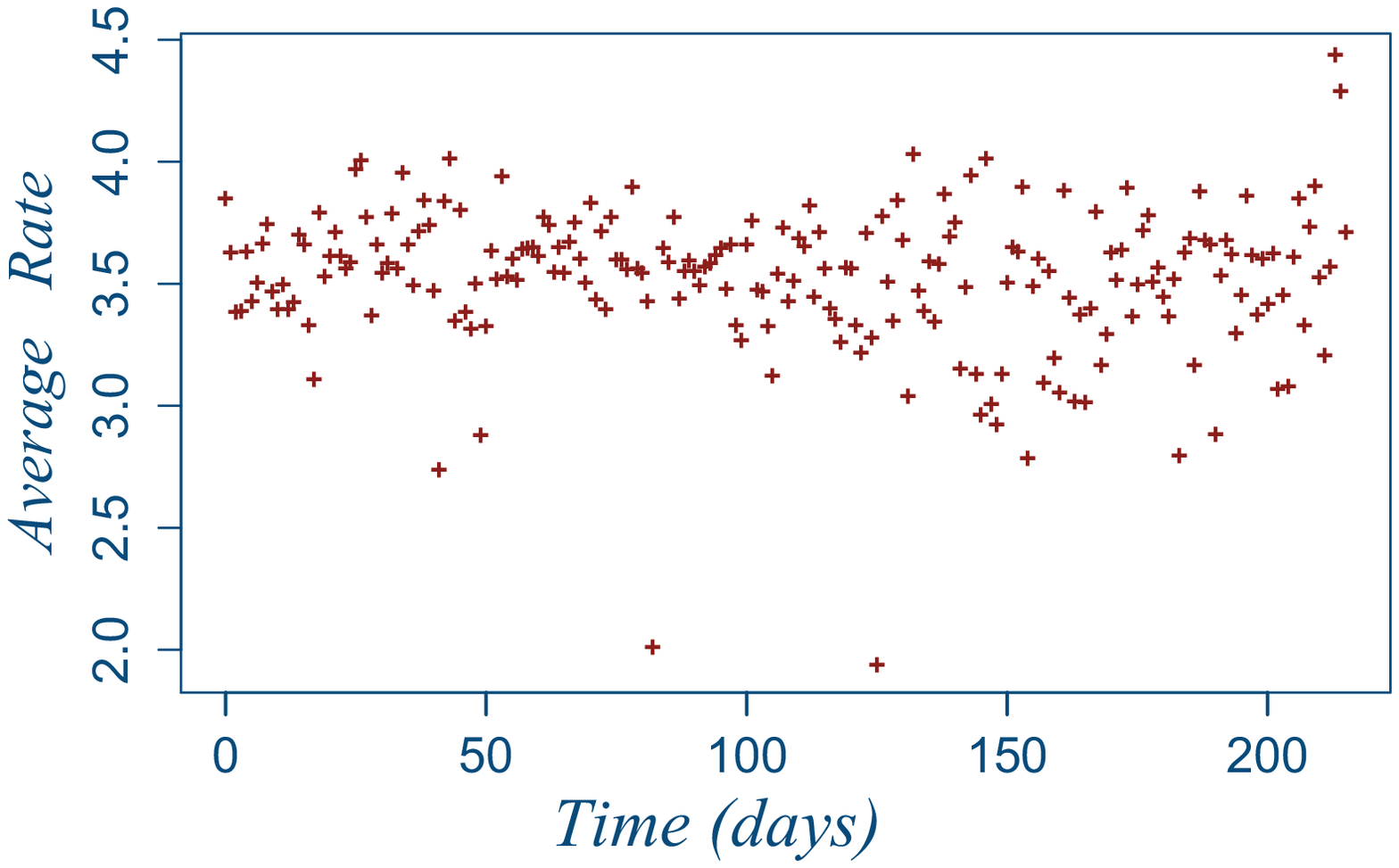}\\
  (a) & (b)
\end{tabular}
\caption{ Temporal effects of the two observed datasets: ML100k and ML1m. $days$ in the x axis indicates the number of days since the first rating in the dataset. (a) depicts the evolution of average movie-rating for $ML1m$, (b) denotes the evolution of average movie-rating for $ML100k$.}
\label{Fig:TAver}
\end{figure*}

In this example, we focus on investigating the day-shifting patterns of customers preferences on movies. Two interesting temporal effects associated with the data are shown in Figure \ref{Fig:TAver}. An significant effect within ML1m is that the mean rating concentrates around 3.5 in the beginning 300 days, but fluctuates in an intensive amplitude later on. The snapshot $S$ for observed data shows that each example contains the time information, which corresponds with a certain day in a year. It's noted that users usually do not insist on logging in the system every day. In order to predict the future changes based on the accumulate limited amount of users' daily behaviors, we should depend on not only users' historical behaviors on old time point, but also the collaborative information from the close neighbors. Thereby time information need to be transformed into a common format. As intending to explore the day-shifting patterns of users, we define a time mapping function $F(t)$, whose output denotes the number of days since the first day of a year. For instance, if input $t$ = $2010$-$01$-$02$, the output of $F(t)$ is 2, which means that input $t$ is associated with the $2nd$ day in 2010. The output of $F(t)$ is independent on a specific year, and only cares about the number of days. In order to model the day-drifting changes, we apply $F$ to the time point for each observed example. Let the modified observed $\bar{S}$ be:
\begin{equation*}
\begin{aligned}
    \bar{S} = \{&(A, Titanic, 2, 5), (A, Star\ Wars, 91, 1),\\
    &(B, Star\ Wars, 124, 4), (C, Time\ Tracers, 91, 4)\}
\end{aligned}
\end{equation*}

Usually, RS could log daily generated data, which ensures that the time factor $t$ can cover all possible values. Future date could find its corresponding feature parameters after being transformed with function $F(t)$. Then using the data like $\bar{S}$ to estimate user $u$'s preferences an uncollected item $i$ at time $t$ can be formulated by:
\begin{equation}
\label{Eq.8}
\widehat{r}_{ui} = \underbrace{p_uq_i^T} _{user\ \#\ item} +\ \underbrace{p_{ut}q_{t}^T } _{user\ \#\ time}.
\end{equation}

\section{Empirical Analysis}
The experiments were conducted on two MovleLens data sets as described in Section 3.2. In the experiments, we applied the 5-fold cross-validation method on both data sets for the first application of MLIMF on extracted features. In order to simulate real recommendation occasion for prediction task on the future changes, we apply all-but-two\footnote{all-but-two: Only the last two ratings of each individual are split into the validation set.} experiment setting for the second application of MLIMF on modeling the time effects. The specific value of temporal recommendation on the evaluation metric means the average results for 5 runs with random initialization.

$Evaluation\ Metric.$ The performance of recommendation algorithms is measured by the root mean squared error (RMSE), a widely used metric for evaluating the rating prediction accuracy of recommenders, given by:
\begin{equation}
    RMSE = \sqrt{ \frac {\sum _{(u,i)\in V}(r_{ui} - \hat{r}_{ui})^2} {|V|} },
\end{equation}
where $V$ denotes the validation set. RMSE measures the errors between the true values and the predictions. Obviously lower RMSE means higher prediction accuracy.

\subsection{Baseline Methods}
In this section, in order to show efficiency of our proposed recommendation method, we compare the recommendation results with the three baseline algorithms.

$User$-$based$ $CF$ (UCF) \cite{lu2012recommender}. UCF is a typical implementation of CF. In UCF, the prediction task for an active user depends on a group of neighbors with similar interests. UCF generates recommendations by two steps: ($i$) calculate the similarity $s_{uv}$, which denotes the correlation or distance between user $u$ and $v$; ($ii$) generate the predictions for an active user by taking the weighted average of all ratings of his/her k Nearest Neighbors ($k$NN). In this paper, the similarity $s_{uv}$ between user $u$ and $v$ is calculated with cosine-based metric. Let $\bar{R}$ denote the set of common items rated by both user $u$ and $v$. Then the similarity is formulated by:
\begin{equation}
sim(u,v) = \frac {\sum _{i\in \bar{R}} r_{ui}r_{vi}} { \sqrt{\sum _{i\in \bar{R}}r_{ui}^2r_{vi}^2} }.
\end{equation}
To predict an active user $u$'s rating on an uncollected item $i$, we can take a weighted average of all the ratings on that item according to the following formula \cite{resnick1994grouplens}:
\begin{equation}
\hat{r}_{ui} = \bar{r}_u + \frac {\sum _{v\in N_u} (r_{vi} - \bar{r}_v)s_{uv}} {\sum _{v\in N_u}s_{uv} },
\end{equation}
where $\bar{r}_u$ and $\bar{r}_v$ denote the mean rating of user $u$ and $v$, respectively $N_u$ denotes the set of user $u$'s nearest neighbors who has collected item $i$.

$Item$-$based$ $CF$ (ICF) \cite{sarwar2001item}. Rather than computing the similarity between user pairs, ICF starts from matching the user's rated items with similar items, then combines the most similar ones into recommendation list. We employ the cosine-based correlation to measure the similarity between item pairs. Let the $C$ denote the set of common users involved with both item $i$ and $j$. Then the similarity is calculated by:
\begin{equation}
sim(i,j) = \frac {\sum _{u\in \bar{C}} r_{ui}r_{uj}} { \sqrt{\sum _{u\in \bar{C}}r_{ui}^2r_{uj}^2} }.
\end{equation}
The prediction step is significant in producing recommendation list. In ICF, generating recommendation results to an active user is based on his/her historical rated items. In this work, the estimate of $r_{ui}$ for active user $u$ is computed by:
\begin{equation}
\hat{r}_{ui} = \bar{r}_i + \frac {\sum _{j\in R_u} (r_{uj} - \bar{r}_j)s_{ij}} {\sum _{j\in R_u}s_{ij} },
\end{equation}
where $R_u$ denotes the set of rated items by user $u$. $\bar{r}_i$ and $\bar{r}_j$ respectively denote the mean rating of item $i$ and $j$.

$RMF$ \cite{webb2006rmf,koren2009matrix}: This method has been described in the Section 2. It uses only the user-rating matrix to generate recommendations.

\subsection{Performance validation for MLIMF}
$Experiment\ settings.$ In this part we intend to validate the performance of our proposed MLIMF with other three baseline algorithms. In order to make MF models converge at the optimized result, the initial values of feature vectors for both RMF and MLIMF are randomly drawn from a normal distribution $\mathcal{N}$(0, 0.02), following the suggestion from \cite{takacs2009scalable}. Note that, the incorporation of extra factors makes  MLIMF more difficulty to set the value of dimension parameter for the comparative experiments on MLIMF and RMF. Generally, MF-based approaches can produce more precise rating estimation with the growth of their feature dimension. Different from RMF, MLIMF has additional feature dimension parameters, i.e. $f_{D_j}$, which makes it difficult to compare their performance as the reason mentioned above. We can not set $f_i$ as the same value of $f$, for it will result in that MLIMF has large value of feature dimension parameter. As the basic MF method, RMF only models the interactions between user-item pairs, which simplifies the procedure of pre-defining the dimension parameter $f$ for user and item feature vectors. Usually, user-item pairs are matched into the same latent feature space with dimension $f$ according to the principles of RMF. $f$ is the key parameter to influence the efficiency of modeling user preferences on items. Therefore, we should redefine the parameter $f$ for MLIMF because MLIMF models the interactions between users and extra factors besides user-item pairs. Since the objective is to validate the performance of the proposed MLIMF, we design the experimental processes as follows:
\begin{itemize}
    \item[-] Firstly, we modify the settings of dimension parameters for RMF and MLIMF. The given value of dimension parameter $f$ for RMF equals to the sum of dimension variables for MLIMF, which can be denoted as:
          \begin{equation}
              f = f_i + \sum _{j\in D}f_{D_j},
          \end{equation}

        where $f_i$ and $f_{D_j}$ are respectively the pre-defined parameters for the dimension of the user-item and user-factor feature spaces. In the application of MLIMF on extracted features, dimension parameter $f_{D_j}$ for each decision factor $j$ equals to 20\% of the given value of $f$, which means $f_i$ = 0.4$*f$ and $f_{D_j}$ = 0.2$*f$. In applying MLIMF to modeling time effects, the only parameter $f_{D_t}$ for decision factors equals to 60\% of the given value of $f$. In this work, that we set the sum of feature dimensions of MLIMF to the same value of $f$ for RMF is purely for equally comparing their performance with the original idea that fixing the value of dimension parameter $f$ of RMF and using Eq. (16) to initialize feature dimensions of MLIMF could help us better compare both MF-based approach. The float values like 0.4, 0.2 are empirically used to express the significance of corresponding attribute for users’ final rating decision. In practice, one can independently allocate a dimensional value to different decision factor according to their own prior knowledge.
       \item[-] After implementing detailed analysis on experiment datasets, we carry on two possible examples to deeply clarify the idea of MLIMF. Furthermore, comparative experiments for those examples are conducted for MLIMF and other three baseline approaches, including UCF, ICF, RMF. The experimental results are depicted in Tab. \ref{tab:template}.
       \item[-] According to \cite{takacs2009scalable}, the dimension of feature space can greatly affect the accuracy of MF-based approaches. Then we explore the impact of different values of dimension parameter $f$ on the accuracy of RMF and MLIMF. The experimental results are shown in Tab. \ref{tab:extracted} and Tab. \ref{tab:temporal}.
\end{itemize}

\begin{figure*}[htb]
\centering
\begin{tabular}{cc}
  \includegraphics[scale = 0.36,natwidth=414,natheight=434]{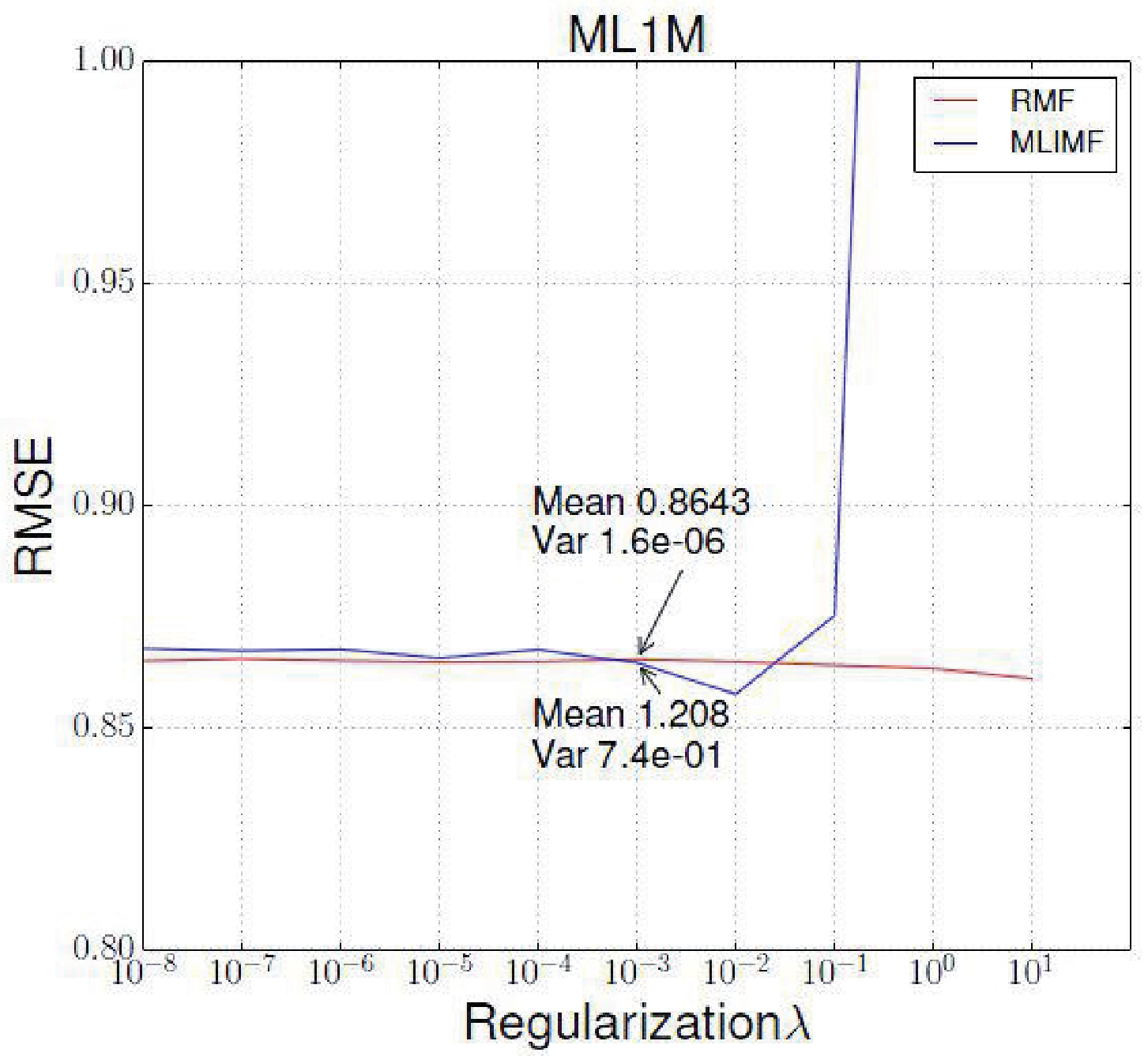}
  &\includegraphics[scale = 0.36,natwidth=414,natheight=433]{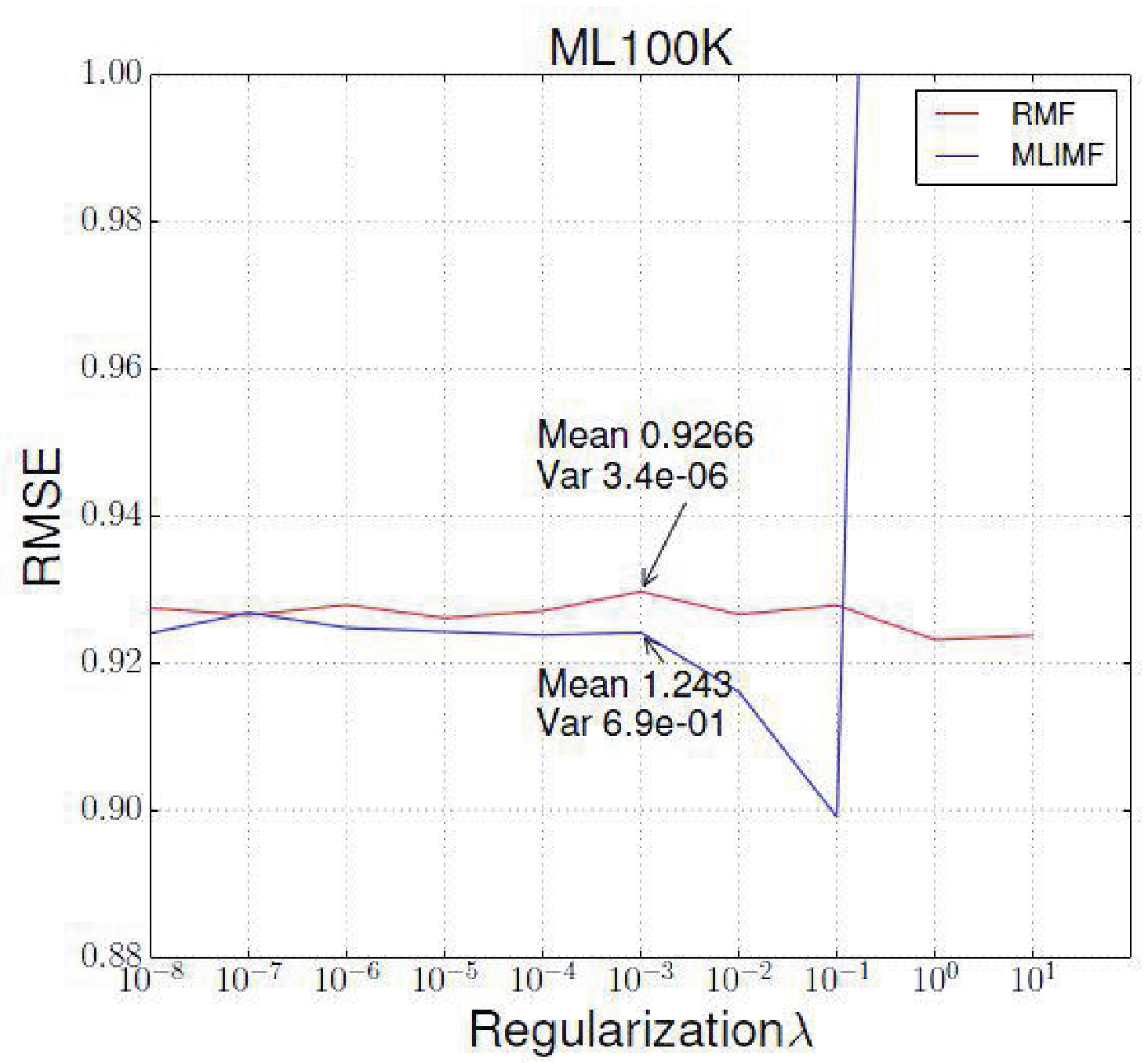}\\
  (a) & (b)
\end{tabular}
\caption{ Illustrating the relevance of selecting different regularization parameters $\lambda$ and the performance of RMF and MLIMF with a fixed setup, where dimension is set to 10, learning ratio is set to 0.01. It ‘s noted that RMSE for MLIMF in two dataset when $\lambda$ equals to 1 or 10, is too large. Therefore, we just do not illustrate their value in order to clearly distinguish the performances of RMF and MLIMF in small $\lambda$.}
\label{Fig:REG}
\end{figure*}

In the experimental processes, all aforementioned methods have many pre-defined parameters that greatly influence the accuracy. For UCF, the number of nearest neighbors $k$ is significant for building UCF with high accuracy. After conducting several experiments to explore the accuracy of UCF, we decide to use the top 25\% neighbors of each user to generate predictive score of an uncollected item. The initial values of all features vectors are randomly chosen from a normal distribution $\mathcal{N}$(0, 0.02). Usually, too small, or large value of $\lambda$ will lead to very low generality on the testing dataset~\cite{bishop2006pattern}. To our knowledge, most of works~\cite{takacs2008investigation, koren2008factorization, luo2012incremental} based on MF method will set $\lambda$ in a comparatively small interval [0.001, 0.1]. To better present the influence of the selection of regularization parameter $\lambda$, we further do some experiments on two movielens datasets. Results are illustrated in Figure \ref{Fig:REG}. For the simiplicity, we conduct experiments in dataset from the first application case (Extracted Features). In this work, for both experiment data sets ML1m and ML100k, the regularizing parameter $\lambda$ for RMF and MLIMF is set to 0.01. In terms of the learning rate for RMF and MLIMF, initially $\gamma$ and $\eta$ are both set to 0.01 to ensure a comparative fast convergence rate for ML100k and ML1m. According to Ref. \cite{takacs2009scalable}, in addition to the setting of learning rate, the optimized result of the objective function $J$ is correlated with the density of user-item rating matrix $R$. Interestingly, in MLIMF the user-factor matrix $R_j$ is always denser than $R$ because in real-world application $|D_j|$ is much less than $|I|$. Based on the above consideration, we decline the value of $\eta$ to slow down the updating amplitude after several epoches.
\vspace{-0.5cm}

\begin{table}[ht] 
\centering 
    \begin{tabular}{l l l l c c c c} 
    \toprule 
    \multirowbt{2}{*}{Cases} & \multirowbt{2}{*}{Data} & \multirowbt{2}{*}{UCF} & \multirowbt{2}{*}{ICF}
    & \multicolumn{2}{c}{RMF} & \multicolumn{2}{c}{MLIMF} \\ 
    \cmidrule{5-6} \cmidrule{7-8}
    &&& & $f=20$ &$f=50$& $f=20$ & $f=50$\\
    \midrule
    \multirowbt{2}{2cm}{Extracted Features} & ML100k & 0.953 & 0.940 & 0.918 & 0.913 & 0.906 & \textbf{0.904}\\
    \cmidrule{2-8} 
     & ML1m & 0.933 & 0.909 & 0.863 & 0.860 & 0.855 & \textbf{0.853}\\
    \midrule
    \multirowbt{2}{2cm}{Temporal Dynamic} & ML100K & 1.057 & 1.034 & 1.015 & 1.013 & 1.006 & \textbf{1.004}\\
    \cmidrule{2-8}
     & ML1m & 0.978 & 0.958 & 0.907 & 0.905 & 0.903 & \textbf{0.902}\\

    \bottomrule 
    \end{tabular}
\caption{Comparisons of four recommendation approaches on two application scenarios of MLIMF.} 
\label{tab:template} 
\end{table}

$Experimental\ Results.$ For a clear view, we summarize the RMSE of all mentioned methods in Tab. \ref{tab:template}, which presents that both two MF methods outperform the UCF and ICF when the value of $f$ is set to 20 and 50 respectively. Tab. \ref{tab:template} also shows that the effects of incorporating extra information make MLIMF work better than RMF with the same value of $f$ on two applications of MLIMF.

\begin{table}[ht] 
\centering 
    \begin{tabular}{l l l r r} 
    \toprule 
    \multirowbt{2}{*}{Dimension} & \multicolumn{2}{c}{ML100k} & \multicolumn{2}{c}{ML1m} \\ 
    \cmidrule{2-3} \cmidrule{4-5}
    & RMF &MLIMF& RMF & MLIMF\\
    \midrule
     $f$ = 50 & 0.9133 & \textbf{0.9035} & 0.8593 & \textbf{0.8530}\\
     $f$ = 100 & 0.9108 & 0.9012 & 0.8564 & 0.8513\\
     $f$ = 200 & 0.9091 & 0.8992 & 0.8545 & 0.8500\\
     $f$ = 300 & 0.9081 & 0.8984 & \textbf{0.8534} & 0.8497\\
     $f$ = 400 & 0.9076 & 0.8979 & 0.8527 & 0.8492\\
     $f$ = 500 & \textbf{0.9070} & 0.8972 & 0.8522 & 0.8491\\
    \bottomrule 
    \end{tabular}
\caption{Comparisons of RMF and MLIMF with various settings of $f$ in case of the extracted features.} 
\label{tab:extracted} 
\end{table}

\subsubsection{Impact of Parameter $f$}
Our model is based on RMF. The pre-defined parameter $f$ plays the key role in affecting the optimized accuracy. In this part the detailed analysis of this effect is shown in Tab. \ref{tab:extracted} and Tab. \ref{tab:temporal}. From Tab. \ref{tab:extracted} and Tab. \ref{tab:temporal} we can observe an evident effect of incorporating the extracted features into MLIMF. With different settings of $f$ on both ML100k and ML1m, MLIMF always yields lower RMSE than RMF model. Moreover, for ML100k we can clearly observe that the MLIMF with $f=50$ could mimic the optimum RMSE produced by RMF with $f=500$. For ML1m, MLIMF with $f=50$ could generate the optimized RMSE yielded by RMF with $f=300$. In particular, results in Tab. \ref{tab:temporal} also indicate the effectiveness of MLIMF on modeling the time effects.

\begin{table}[ht] 
\centering 
    \begin{tabular}{l c c c c} 
    \toprule 
    \multirowbt{2}{*}{Dimension} & \multicolumn{2}{c}{ML100k} & \multicolumn{2}{c}{ML1m} \\ 
    \cmidrule{2-3} \cmidrule{4-5}
    & RMF &MLIMF& RMF & MLIMF\\
    \midrule
         $f$ = 50 & 1.0131 & \textbf{1.0042} & 0.9047 & \textbf{0.9021}\\
         $f$ = 100 & 1.0113 & 1.0022 & 0.9038 & 0.9008\\
         $f$ = 200 & 1.0105 & 1.0012 & 0.9024 & 0.8999\\
         $f$ = 300 & 1.0098 & 1.0007 & \textbf{0.9020} & 0.8993\\
         $f$ = 400 & 1.0096 & 0.9996 & 0.9017 & 0.8990\\
         $f$ = 500 & \textbf{1.0093} & 0.9992 & 0.9016 & 0.8986\\
    \bottomrule 
    \end{tabular}
\caption{Comparisons of RMF and MLIMF with various values of $f$ in case of the time effects. In this case, ``one day" is the time unit.} 
\label{tab:temporal} 
\end{table}

\subsection{Complexity Analysis}
Based on the aforementioned RMSE comparison results, it can be seen that the MLIFM is capable to extend RMF by incorporating additional information. In terms of the training computation complexity, each training epoch, with regard to the sample with format $[u$, $i$, $d_1$, $...$, $d_{|D|}]$, is associated with the following operations:
\begin{itemize}
    \item[-] updating all the latent user features $p_u$ and $p_{ud_j}$ under the rules of Eq. (\ref{EUpdating}), which results in computational complexity at $O(|T|\cdot (f_i + \sum _{j\in D} f_{D_j}) )$. $|T|$ denotes the size of training sample.
    \item[-] updating the latent features for items in user $u$'s rating set, which takes a computational complexity at $O(|T| \cdot f_i)$.
    \item[-] updating the latent features for different factors corresponding to user $u$'s rating records, which totally costs a computational complexity at $O(|T|\cdot \sum _{j\in D} f_{D_j})$.
\end{itemize}
Since above updating steps could be done within the same iteration, the worst computation complexity of MLIMF on modeling interesting patterns of the data is $O( (|T| + |T| + |T| ) \cdot f')$, where $f'$ equals to $f_i + \sum _{j\in D} f_{D_j}$. Given the iterative times $n$ for convergence, then the computation complexity for updating over $r_{ui}$ in MLIMF can be formulated by:
\begin{equation}
O\left(|T| \times n \times f'\right),
\end{equation}
which depicts a fact that though taking into account extra factors, the computation complexity of MLIMF grows in linear time comparing with RMF. In terms of the space complexity, it increases with the number of factors incorporated into the proposed MLIMF model.

\section{Conclusions and Discussion}
Many pioneering researches have proved that Matrix Factorization (MF) based approaches are effective and flexible in dealing with various aspects of user-item rating data. Generally, the final rating decisions of online users should be affected by various underlying factors, such as emotions, time, genres, and so on. In this paper, based on classical MF method, we propose a multi-linear interactive MF (MLIMF) approach, trying to gain insight into user preferences. Firstly, we assume that users are willing to implicitly or explicitly weigh the impact of each factor when they rate items. Secondly, we extract possible factors correlated with users' decisions from empirical analyses. Thirdly, to model the multiple pairwise relationship, we linearly integrate the total pairwise interactions to predict their ratings. Finally, experiments results show that the proposed MLIMF method perform much better than three baseline algorithms (UCF, ICF and RMF) with the RMSE metric. 

Overall, MLIMF is a simple yet general approach since it mainly focuses on modeling the interactions between user and other information beyond ratings. Similar inspiration can be easily applied to other MF based models as a bulk denoting the user-factor interactions. In this paper, we just simply extend the basic RMF to explore the impact of categorical attributes on users' rating patterns. However, there are many data mining tasks which need to deal with attributes of continuous values. Therefore, in order to address more general data mining challenges, it is necessary to design an effective framework to extend the proposed MLIMF. We attempt to study the possible applications of MLIMF to solve tough recommendation challenges like estimating click-through rate (CTR) in the era of computational advertising, building effective binary classifiers to predict the potential tastes of online users.

\section{Acknowledgements}
This work was partially supported by the National Natural Science Foundation of China (Grant Nos. 11305043 and 11301490), the EU FP7 Grant 611272 (project GROWTHCOM), Zhejiang Provincial Natural Science Foundation of China (Grant No. LY14A050001), and the Zhejiang Provincial Qianjiang Talents Project (Grant No. QJC1302001), the start-up foundations of Hangzhou Normal University.





\bibliographystyle{elsarticle-num}
\bibliography{kbs}

\end{document}